\documentclass[twoside,twocolumn,9pt]{article}
\usepackage{extsizes}
\usepackage[super,sort&compress,comma]{natbib} 
\usepackage[version=3]{mhchem}
\usepackage[left=1.5cm, right=1.5cm, top=1.785cm, bottom=2.0cm]{geometry}
\usepackage[dvipsnames]{xcolor}
\usepackage{balance}
\usepackage{mathptmx}
\usepackage{sectsty}
\usepackage{graphicx}
\graphicspath{{figures/}}
\usepackage{subcaption}
\usepackage{lastpage}
\usepackage[
  format=plain,
  justification=justified,
  singlelinecheck=false,
  font={stretch=1.125,small,sf},
  labelfont=bf,
  labelsep=space
]{caption}
\usepackage{float}
\usepackage{fancyhdr}
\usepackage{comment}
\usepackage{fnpos}
\usepackage[english]{babel}
\addto{\captionsenglish}{%
  
}
\usepackage{array}
\usepackage{droidsans}
\usepackage{charter}
\usepackage[T1]{fontenc}
\usepackage{setspace}
\usepackage[compact]{titlesec}
\usepackage{hyperref}
\usepackage{amssymb}
\usepackage{epstopdf}
\usepackage[normalem]{ulem}
\definecolor{darkBlue}{rgb}{0,0,0.6}
\definecolor{cream}{RGB}{222,217,201}

\makeatletter
\newcommand{\smartfootnotemark}[1]{%
  \begingroup
    \ifnum#1<6
      \renewcommand{\thefootnote}{${\mathit{\alph{footnote}}}$}%
      \footnotemark[#1]%
    \else
      \renewcommand{\thefootnote}{\fnsymbol{footnote}}%
      \footnotemark[\numexpr#1-4\relax]%
    \fi
  \endgroup
}
\newcommand{\smartfootnotetext}[2]{%
  \begingroup
    \ifnum#1<6
      \renewcommand{\thefootnote}{$^{\displaystyle\mathit{\alph{footnote}}}$}%
      \footnotetext[#1]{#2}%
    \else
      \renewcommand{\thefootnote}{\fnsymbol{footnote}}%
      \footnotetext[\numexpr#1-4\relax]{#2}%
    \fi
  \endgroup
}
\makeatother

\begin{document}

\pagestyle{fancy}

\thispagestyle{plain}
\fancypagestyle{plain}{
%%%HEADER%%%
\renewcommand{\headrulewidth}{0pt}
}
%%%END OF HEADER%%%

%%%PAGE SETUP - Please do not change any commands within this section%%%
\makeFNbottom
\makeatletter
\renewcommand\LARGE{\@setfontsize\LARGE{15pt}{17}}
\renewcommand\Large{\@setfontsize\Large{12pt}{14}}
\renewcommand\large{\@setfontsize\large{10pt}{12}}
\renewcommand\footnotesize{\@setfontsize\footnotesize{7pt}{10}}
\makeatother

\renewcommand{\thefootnote}{\fnsymbol{footnote}}
\renewcommand\footnoterule{\vspace*{1pt}% 
\color{cream}\hrule width 3.5in height 0.4pt \color{black}\vspace*{5pt}} 
\setcounter{secnumdepth}{5}

\makeatletter 
\renewcommand\@biblabel[1]{#1}            
\renewcommand\@makefntext[1]% 
{\noindent\makebox[0pt][r]{\@thefnmark\,}#1}
\makeatother 
\renewcommand{\figurename}{\small{Fig.}}
\sectionfont{\sffamily\Large}
\subsectionfont{\normalsize}
\subsubsectionfont{\bf}
\setstretch{1.125} %In particular, please do not alter this line.
\setlength{\skip\footins}{0.8cm}
\setlength{\footnotesep}{0.25cm}
\setlength{\jot}{10pt}
\titlespacing*{\section}{0pt}{4pt}{4pt}
\titlespacing*{\subsection}{0pt}{15pt}{1pt}
%%%END OF PAGE SETUP%%%

%%%FOOTER%%%
\fancyfoot{}
\fancyfoot[LO,RE]{\vspace{-7.1pt}\includegraphics[height=9pt]{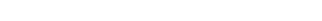}}
\fancyfoot[CO]{\vspace{-7.1pt}\hspace{13.2cm}\includegraphics{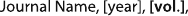}}
\fancyfoot[CE]{\vspace{-7.2pt}\hspace{-14.2cm}\includegraphics{head_foot/RF}}
\fancyfoot[RO]{\footnotesize{\sffamily{1--\pageref{LastPage} ~\textbar  \hspace{2pt}\thepage}}}
\fancyfoot[LE]{\footnotesize{\sffamily{\thepage~\textbar\hspace{3.45cm} 1--\pageref{LastPage}}}}
\fancyhead{}
\renewcommand{\headrulewidth}{0pt} 
\renewcommand{\footrulewidth}{0pt}
\setlength{\arrayrulewidth}{1pt}
\setlength{\columnsep}{6.5mm}
\setlength\bibsep{1pt}
%%%END OF FOOTER%%%

%%%FIGURE SETUP - please do not change any commands within this section%%%
\makeatletter 
\newlength{\figrulesep} 
\setlength{\figrulesep}{0.5\textfloatsep} 

\newcommand{\topfigrule}{\vspace*{-1pt}% 
\noindent{\color{cream}\rule[-\figrulesep]{\columnwidth}{1.5pt}} }

\newcommand{\botfigrule}{\vspace*{-2pt}% 
\noindent{\color{cream}\rule[\figrulesep]{\columnwidth}{1.5pt}} }

\newcommand{\dblfigrule}{\vspace*{-1pt}% 
\noindent{\color{cream}\rule[-\figrulesep]{\textwidth}{1.5pt}} }

\makeatother
%%%END OF FIGURE SETUP%%%

%%%TITLE, AUTHORS AND ABSTRACT%%%
\twocolumn[
  \begin{@twocolumnfalse}
{\includegraphics[height=30pt]{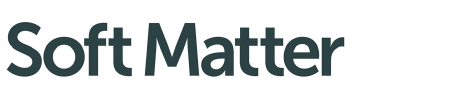}\hfill\raisebox{0pt}[0pt][0pt]{\includegraphics[height=55pt]{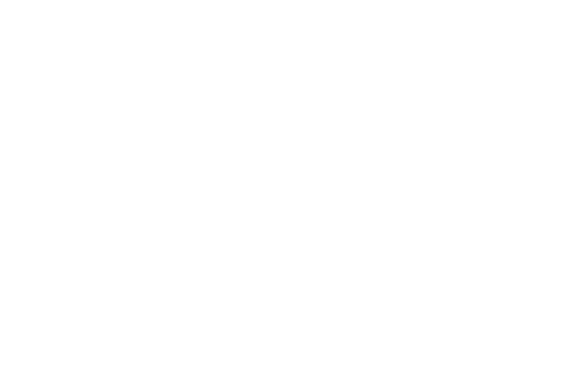}}\\[1ex]
\includegraphics[width=18.5cm]{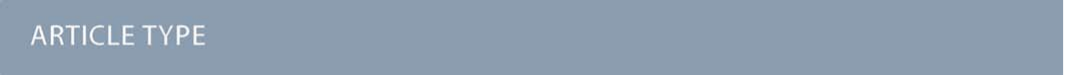}}\par
\vspace{1em}
\sffamily
\begin{tabular}{m{4.5cm} p{13.5cm} }

\includegraphics{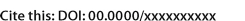} & \noindent\LARGE{\textbf{Transport Properties of Active Particles Moving on Adjustable Networks%$^\dag$
}} \\%Article title goes here instead of the text "This is the title"
\vspace{0.3cm} & \vspace{0.3cm} \\

 & \noindent\large{
 
William G.~C.~Oropesa,\smartfootnotemark{6}\smartfootnotemark{1}\smartfootnotemark{2} Pablo de Castro,\smartfootnotemark{1}\smartfootnotemark{2}\smartfootnotemark{3} Hartmut Löwen,\smartfootnotemark{4} and Danilo B.~Liarte\smartfootnotemark{7}\smartfootnotemark{1}\smartfootnotemark{2}\smartfootnotemark{5}
 
 }\\%Author names go here instead of "Full name", etc.

\includegraphics{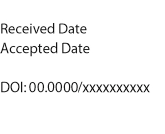} & \noindent\normalsize{%
Active adaptive matter has attracted considerable interest due to its rich, largely unexplained dynamics and its relevance to a wide range of synthetic and biological materials.
An important subclass of such systems consists of active particles that can remodel the network in which they move.
Here, we introduce a minimal yet versatile model of active particles moving on an adjustable network.
In this model, particles undergo discrete run-and-tumble motion along the links of a triangular lattice and leave behind a trail of temporarily blocked links.
These closed links cannot be traversed by other particles and reopen only after a characteristic healing time.
The resulting trail-mediated blocking mechanism is fundamentally distinct from more familiar interactions such as excluded-volume effects.
In the high-persistence limit, we find a qualitative contrast between the two mechanisms: while steric blocking leads to reduced diffusivity with increasing persistence, trail-induced blocking causes diffusivity to increase monotonically.
We characterize this fundamental difference and the unexpected transport properties that arise when both blocking mechanisms are present, and discuss potential applications.
%The abstract should be a single paragraph which summarises the content of the article. Any references in the abstract should be written out in full \textit{e.g.}\ [Surname \textit{et al., Journal Title}, 2000, \textbf{35}, 3523].
} \\%The abstrast goes here instead of the text "The abstract should be..."

\end{tabular}

 \end{@twocolumnfalse} \vspace{0.6cm}

  ]
%%%END OF TITLE, AUTHORS AND ABSTRACT%%%

%%%FONT SETUP - please do not change any commands within this section
\renewcommand*\rmdefault{bch}\normalfont\upshape
\rmfamily
\section*{}
\vspace{-1cm}

%%%FOOTNOTES%%%

%%%FOOTNOTES%%%

\smartfootnotetext{1}{Instituto de Física Teórica, UNESP -- Universidade Estadual Paulista, Rua Dr.~Bento T.~Ferraz 271, 01140-070, São Paulo, SP, Brazil}
\smartfootnotetext{2}{ICTP South American Institute for Fundamental Research, São Paulo, SP, Brazil}
\smartfootnotetext{3}{Instituto de Física, Universidade de São Paulo, 05508-090 São Paulo, São Paulo, Brazil}
\smartfootnotetext{4}{Institut für Theoretische Physik II: Weiche Materie, Heinrich-Heine-Universität Düsseldorf, Universitätsstr. 1, D-40225 Düsseldorf, Germany}
\smartfootnotetext{5}{Department of Physics, Cornell University, Ithaca, NY 14853, USA}
\smartfootnotetext{6}{~E-mail: oropesaw@ictp-saifr.org}
\smartfootnotetext{7}{~E-mail: danilo.liarte@ictp-saifr.org}

%Please use \dag to cite the ESI in the main text of the article.
%If you article does not have ESI please remove the the \dag symbol from the title and the footnotetext below.
%\footnotetext{\dag~Supplementary Information available: [details of any supplementary information available should be included here]. See DOI: 10.1039/cXsm000v00x/}
%additional addresses can be cited as above using the lower-case letters, c, d, e... If all authors are from the same address, no letter is required

%%%END OF FOOTNOTES%%%

%%%MAIN TEXT%%%%
\section{Introduction}
\label{sec:Introduction}

In recent years, a wide range of striking phenomena have been reported in connection with the unusual transport properties of active matter~\cite{ramaswamy2010mechanics, marchetti2013hydrodynamics, elgeti2015physics, bechinger2016active, te2025metareview, gompper20202020, vrugt2025exactlyactivematter, mestre2022colloidal,villa2020run}.
Many of these effects can be traced back to what is arguably the most ubiquitous feature of active matter systems: \emph{persistent motion}.
As a consequence of persistence, a single active particle undergoing stochastic motion remains in a super-diffusive regime for times up to the \emph{persistence time}, defined as the average time over which the particle maintains its direction of motion~\cite{escobar2025effect,agoritsas2024memoryformationdensepersistent}.
In interacting systems, persistence gives rise to even more remarkable behavior.
Notably, it can drive spontaneous \emph{agglomeration} or \emph{clustering} even in systems with purely repulsive interactions~\cite{WensinkLoewen2008, theurkauff2012dynamic, caprini2024dynamical, levay2025cluster}.
In its most prominent manifestation, this mechanism leads to the well-known phenomenon of \emph{motility-induced phase separation} (MIPS)~\cite{cates2015motility, palacci2013living, PhysRevLett.110.238301,maggi2021universality,martin2025motility,omar2023mechanical}.
Persistence also plays a crucial role in determining long-time transport properties.
For instance, the diffusion coefficient of an active particle moving in a porous medium exhibits a nonmonotonic dependence on the persistence time: in the large-persistence limit, diffusion is suppressed due to persistence-induced accumulation at confining boundaries~\cite{kurzthaler2021geometric, debets2021cage}.
Similarly, in systems of mutually repulsive active particles, the collective diffusion coefficient also displays a nonmonotonic dependence on persistence, primarily as a result of clustering~\cite{MartensBocquet2012,LevisBerthier2014,SotoGolestanian2014}.
Here we consider a more general setting in which particle motion actively induces topological changes in the embedding medium --- in our case a discrete network.
We show that this coupling leads to a novel mechanism of diffusion suppression that is fundamentally distinct from the interaction-driven slowing down associated with agglomeration
\cite{solon2015flocking, scandolo2023active, bandyopadhyay2024ordering, dittrich2021critical, loos2023long}.

Network topology adjustment has emerged as a key mechanism underlying a wide range of phenomena in both synthetic~\cite{jin2017chemotaxis} and biological materials.
In vertex models of confluent cell tissues, changes in network topology provide the essential mechanism that enables cell rearrangements and mobility in the fluid-like regime~\cite{BiManning2015,BiManning2016}.
More generally, systematic bond removal in elastic networks—known as the tuning-by-pruning protocol—has been shown to regulate and optimize the mechanical response of diverse material systems~\cite{GoodrichNagel2015,ReidPablo2018,LiarteLubesnky2020}.
This framework was subsequently extended to achieve localized, long-range, and controllable responses in elastic and flow networks~\cite{RocksNagel2017,RocksKatifori2019}.
In this mechanical analogue of allosteric regulation, a localized strain (or pressure drop) applied at position $\boldsymbol{r}$ induces a prescribed, spatially separated response at position $\boldsymbol{r}^\prime$.
More recently, networks with adaptable degrees of freedom have been equipped with local learning rules that enable them to modify their internal structure and optimize performance in response to sequences of external stimuli~\cite{SternLiu2021,HaghCorwin2022,DillavouDurian2024,HainManning2025}.
This line of work has employed the paradigm of physical learning~\cite{SternMurugan2023}, which has generated significant interest as a framework for understanding complex adaptive systems and for designing intelligent mechanical metamaterials.

Recently, there has been growing interest in the study of active particles moving within adaptable and dynamic media.
For example, Jin et al.\ investigated the dynamics of self-propelled droplet swimmers that exhibit both chemotaxis and negative autochemotaxis~\cite{jin2017chemotaxis}.
As these particles move, they deposit chemical trails that can either attract or repel other particles.
These trails have a finite lifetime due to dispersion.
Using microfluidic devices with bifurcating channels, the authors demonstrated anti-correlations in the branch choices of successive swimmers and showed how chemotactic responses can guide swimmers through a maze.
Another example of the important interplay between active motion and the surrounding medium is provided by the migration of certain unicellular organisms, such as the slime mold \emph{Physarum polycephalum}, on different substrates.
Tr\"oger et al.\ showed that, regardless of the presence or absence of food, these organisms achieve superdiffusive migration by performing self-avoiding run-and-tumble dynamics with trail memory~\cite{TroegerAlim2024}.
Another interesting example of the important role played by memory effects on transport properties of active systems is a recent framework connecting self-diffusion in interacting active matter to collisional forces and their temporal correlations.
Akintunde and Mallory show that in active fluids of repulsive active Brownian particles, transport is controlled by both instantaneous collisional hindrance and memory effects arising from persistent force correlations, with the latter providing a strictly dissipative correction to diffusion~\cite{AkintundeMallory2026}.
Finally, a further system of interest involves fibroblasts and other cell types migrating through complex elastic gels,\cite{goswami2024anomalous,nabizadeh2024network} such as the connective tissue of cartilage~\cite{SilverbergCohen2014,JacksonCohen2022}.
As part of their role in tissue repair, fibroblasts are responsible for the production and maintenance of the extracellular matrix.

Here we introduce a minimal yet versatile model that explicitly couples active particle dynamics to the adaptive evolution of network topology.
Specifically, we consider a system of active particles at packing fraction $\phi$ moving along the bonds of a regular triangular lattice~\footnote{We have checked that alternative lattice geometries, including square and honeycomb lattices, yield the same trends for how transport properties change with our parameters.}. Various phenomena in active matter have been described using a fixed underlying lattice~\cite{loos2023long, bandini2025xy, avni2025nonreciprocal, martin2025transition, khasseh2025active,soto2014run,de2021diversity}, but in our model the lattice topology changes dynamically.
Particle motion follows a discrete run-and-tumble dynamics characterized by a persistence time $\tau_p$.
Crucially, particle trajectories actively modify the underlying network: when a particle traverses a bond $b$ connecting two sites, that bond temporarily closes and becomes inaccessible to all particles, including the particle that created the closure.
Each closed bond reopens after a characteristic healing time $\tau_h$.
Hence the particle motion occurs on an \emph{adjustable network}.
In addition to this trail-mediated interaction, particles experience excluded-volume constraints, such that no two particles may occupy the same lattice site.
We demonstrate that these two blocking mechanisms—self-generated bond closures and mutual excluded-volume interactions—give rise to qualitatively distinct transport regimes.
At long persistence times, excluded-volume interactions combined with directional persistence promote particle agglomeration, leading to a suppression of effective diffusion as persistence increases.
By contrast, transport hindered by closed bonds exhibits the opposite trend: increasing persistence systematically enhances diffusion.
The competition between these two fundamentally different blocking mechanisms underlies the nontrivial and counterintuitive dependence of transport properties on system parameters.
In particular, we find that the optimal persistence time that maximizes diffusion (as a function of $\tau_p$) decreases with increasing packing fraction, yet increases with the bond healing time.
These results highlight how feedback between active motion and adjustable connectivity can qualitatively reshape transport in active systems.

The remainder of the paper is organized as follows.
Sections~\ref{sec:Model} and~\ref{sec:Simulations} describe the model and the numerical simulation methods, respectively.
Section~\ref{sec:Results} presents the results for both fixed network topologies (Section~\ref{subsec:topFixed}) and adjustable network topologies (Section~\ref{subsec:AdjustableNetworks}).
Concluding remarks are provided in Section~\ref{sec:Conclusions}.
Finally, Appendix~\ref{app:PowerLaws} brings an argument for the observed power law for the diffusivity in the long-persistence limit and Appendix~\ref{app:NoParticleInt} examines a variant of the model without excluded-volume interactions, in which multiple particles may occupy the same site.

\section{Network model}
\label{sec:Model}
\begin{figure*}
\begin{center}
  \includegraphics[width=0.9\textwidth]{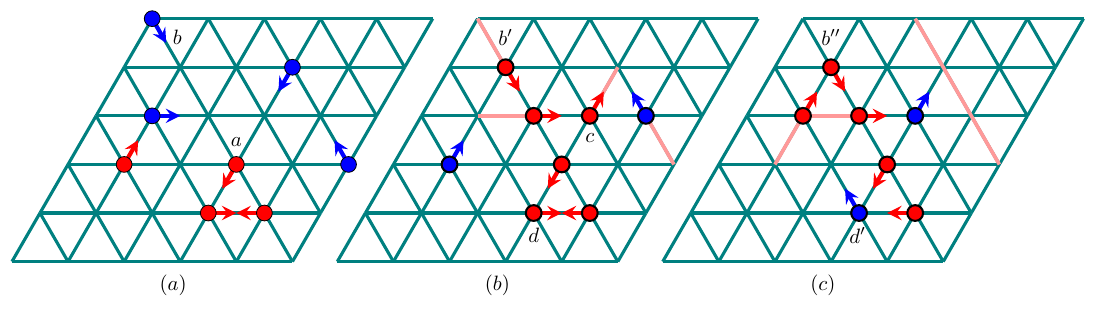}
\end{center}
  \caption{{\bf Illustration of our network model.}
  Run-and-tumble active particles are added to the nodes of a regular triangular lattice with nearest-neighbor connections.
  Blocked and moving particles are represented by red and blue disks, respectively.
  The arrows indicate the direction of self-propulsion.
  Teal and light red lines represent open and closed bonds, respectively. 
  A particle $d$ in (b) changes its configuration to $d^\prime$ in (c) by tumbling, i.e.\ changing its self-propulsion direction at a rate $\alpha_{p} = 1 / \tau_p$.
  An open bond $b$ in (a) traversed by a particle has its configuration changed to a closed bond $b^\prime$ in (b).
  The closed bond $b^\prime$ in (b) can then spontaneously ``heal'' to an open bond state $b^{\prime \prime}$ in (c) at healing rate $\alpha_{h} = 1 / \tau_h$.
  A particle will be blocked either if it tries to move to an occupied site [see particle $a$ in (a)] or if it tries to move through a closed bond [see particle $c$ in (b)].
  }
  \label{fig:Model}
\end{figure*}

We consider a regular triangular lattice with $N$ sites connected by nearest-neighbor bonds between pairs (dark green lines in Fig.~\ref{fig:Model}a).
Each lattice site $i \in \{0,\, \cdots, \, N-1\}$ has nearest neighbors located a lattice spacing away along $z$ possible directions, which we label $\ell \in\{0,\,\dots,\,z-1\}$ ($z=6$ for the triangular lattice).
For future reference, we introduce a lattice index neighbor function $\varphi(i,\ell)\in\{0,\dots,N-1\}$, which indexes the nearest neighbor of $i$ in the direction $\ell$.
In our model, each bond connecting nearest-neighbor sites can be either open or closed.
This is described by a lattice degree of freedom $\kappa_{i \,  \ell} \in \{0, \, 1\}$, which takes value 0 or 1 for closed and open bonds, respectively.
Notice that this variable is redundant: For each site $i$ and direction $\ell$, there is a neighboring site $j=\varphi(i,\, \ell)$ so that $\kappa_{j\, \bar{\ell}} = \kappa_{i \,  \ell}$, where $\bar{\ell}=(\ell+\lfloor z/2\rfloor)\bmod z$ is the opposite direction to $\ell$ and $\lfloor\cdot\rfloor$ is the floor function.

Now we populate this lattice with a set of $N_p$ ($\leq N$) \emph{active} particles, so that each lattice site can be either empty or occupied by at most one particle.
The state of site $i$ is thus described by the occupation variable $\sigma_i\in\{0,1\}$, where 0 and 1 denote empty and occupied sites, respectively.
Alternatively, a configuration of active particles is described by two variables: (i) the coordinate $x_\alpha \in \{0, \, \cdots, \, N-1\}$ giving the lattice site index for particle $\alpha \in \{0,\, \cdots, \, N_p-1\}$ (henceforth we use Greek and Latin indexes for particle and lattice indexes, respectively); (ii) the direction of self-propulsion $n_\alpha \in \{0,\, \cdots ,\, z-1\}$ of particle $\alpha$ (represented by arrows in Fig.~\ref{fig:Model}).
A general configuration for the system is completely determined by the variables $\kappa_{i\, \ell}$, $x_\alpha$ and $n_\alpha$, with $i \in \{1, \, \cdots, \, N-1\}$, $\ell \in \{0, \, \cdots, \, z-1\}$ and $\alpha \in \{0,\, \cdots,\, N_p-1\}$.
In this paper we consider periodic boundary conditions (PBCs).

The particles undergo a run-and-tumble dynamics, i.e., unless a particle $\alpha$ is blocked, it will move along its internal self-propulsion direction (or director) $n_\alpha$.
In turn, this direction can change at each time step $\Delta t$ with probability $\alpha_{p} \Delta t$, where $\alpha_{p}$ is the tumbling rate.
The tumbling rate determines a characteristic time scale, the persistence time $\tau_p = 1 / \alpha_{p}$, giving the average time between consecutive tumbles.
Note that for $\tau_{p}=\Delta t$, we recover the passive limit, where particles follow a traditional dynamics in which the director virtually changes at every time step.
Conversely, for $\tau_{p}\to\infty$ we recover deterministic ballistic motion along a fixed direction with no tumbles.
The tumble process is schematically represented for a particle undergoing the change $d \rightarrow d^\prime$ in Figs.~\ref{fig:Model}b and~\ref{fig:Model}c.

Whenever a particle traverses a bond $\kappa_{i\nu}$, it becomes closed, i.e.\ $\kappa_{i\nu} \rightarrow 0$ (see bonds $b$ and $b^\prime$ in Figs.~\ref{fig:Model}a and~\ref{fig:Model}b).
Once closed, the bond can spontaneously ``heal'' and return to the open state ($\kappa_{i\nu} = 1$) at each time step with probability $\alpha_{h} \, \Delta t$, where $\alpha_{h}$ is the healing rate (see bonds $b^\prime$ and $b^{\prime \prime}$ in Figs.~\ref{fig:Model}b and~\ref{fig:Model}c).
The healing rate determines another important characteristic time scale, the healing time $\tau_h = 1 / \alpha_{h}$, which represents the average time for a closed bond to regenerate.
The limit $\tau_{h}=\Delta t$ represents the case where bonds heal at every time step with certainty, effectively making the system behave as if all bonds were always open.
Finally, a particle $\alpha$ with director $n_\alpha$ at site $x_\alpha$ will be blocked (i.e.\ not move) either if the intended neighboring site is occupied [$\sigma_{\varphi({x_\alpha,} n_\alpha)}=1$; see particle $a$ in Fig.~\ref{fig:Model}a] or if the intended crossing bond is closed [$\kappa_{x_\alpha\, n_\alpha}=0$; see particle $c$ in Fig.~\ref{fig:Model}b].

The dynamic behavior of our system is controlled by three parameters: The persistence time $\tau_{p}$, the healing time $\tau_{h}$ and the packing fraction of active particles $\phi=N_{p}/N$.
Without loss of generality, we consider the self-propulsion speed of the particles $u=a/\Delta t$ to be constant, where $a$ is the lattice spacing.

\section{Numerical Simulations}
\label{sec:Simulations}

In this section, we present details of our large-scale Monte Carlo simulations on modern GPUs.
We consider triangular lattices of size $N=L\times L$, and PBCs. In all simulations the lattice size is kept fixed, independently of the packing fraction $\phi$.
The number of particles is therefore given by $N_{p}=\lfloor \phi N\rfloor$, ensuring that $N_p$ is an integer.
All simulations were implemented in CUDA~\cite{sanders2010cuda,NVIDIA2026}, enabling efficient parallel updates of particle motion and bond dynamics.
Here, particle and lattice topology are simultaneously updated on the GPU, which allows us to simulate large lattices over long timescales with minimal computational overhead.

In all simulations, the initial network configuration is a complete triangular lattice, i.e.\ all bonds are open---$\kappa_{i\ell}(0)=1$ for all $i,\ell$.
We then assign random values for the position and the director of each active particle $\alpha$, which we draw from uniform distributions over the discrete sets of lattice sites and orientations, respectively.
We have verified that, for all regimes investigated, different realizations of the random initial conditions converge to the same long-time behavior.
In particular, the system consistently reaches a diffusive regime displaying the same qualitative and quantitative characteristics, with no dependence on the initial configuration beyond the expected statistical fluctuations.
At each Monte Carlo step (MCS), the dynamics follows a two-stage protocol:
\begin{enumerate}
    \item \textit{Particle motion.} We update the state of each particle $\alpha$ with director $n_{\alpha}$ at site $i$ in two steps.
    (a) The director changes with probability $\Delta t / \tau_p$ to a new direction uniformly distributed among the discrete $z$ lattice directions.
    (b) The particle attempts to move to its neighboring site $j=\varphi(i,n_{\alpha})$ along the direction of its current director.
    The move is accepted if the target site is empty ($\sigma_j = 0$) and the bond connecting the two sites is open ($\kappa_{i n_{\alpha}} = 1$).
    If the move is successful, we exchange the occupancy variables $\sigma_{i}$ and $\sigma_{j}$, and update the coordinate $x_{\alpha}$, as well as the bond variable $\kappa_{i n_{\alpha}} = 0$, indicating that that bond is now closed.
   \item \textit{Bond ``healing''.}
   After all particle updates, each closed bond is allowed to `heal'.
   A closed bond will be open again with probability $\Delta t / \tau_h$.
   If the healing is successful, we update the bond variable $\kappa_{i\ell} = \kappa_{j\overline{\ell}} = 1$ for $j=\varphi(i,\ell)$.
\end{enumerate}
For each triplet of parameters $(\phi,\tau_p,\tau_h)$, we first evolve the system for a total of $N_{\mathrm{ther}}=10^{5}$ MCS.
The onset of the steady-state regime is determined from the time evolution of the mean squared displacement (MSD), which displays a clear diffusive scaling, $\mathrm{MSD}(t)\sim t$, for $t \gtrsim 10^{4}$ MCS. All observables are subsequently measured within the time window $10^{4} \leq t \leq 10^{5}$ MCS, where the dynamics is stationary.
We employ independent realizations initialized with distinct random-number seeds to compute ensemble averages and estimate statistical uncertainties.
The number of realizations depends on the packing fraction.
For $\phi<0.128$, we use 10 independent realizations, whereas for $\phi\geq0.128$, we use 5 realizations.
We have verified that these numbers of realizations are sufficient and lead to statistically stable averages; see Appendix \ref{app:StatConv} for details above convergence.

%\subsection{Parameters and finite-size considerations}
We have used triangular lattices of linear size $L = 1024$, and the particle packing fraction was varied in the range $0.016 \leqslant \phi \leqslant 0.256$, which is well below the triangular-lattice site-percolation threshold $\phi_c = 1/2$~\cite{sykes1964exact}. The number of particles used varies between $N_{p}\approx 1.7\times 10^{4}$ and $N_{p}\approx 3\times10^{5}$ as $\phi$ is varied for fixed $L$.
We also fixed the units for all simulations: the time step was set to $\Delta t = 1$ and the lattice constant to $a = 1$.
Henceforth, unless stated otherwise, all times (including $\tau_h$ and $\tau_p$) are expressed in units of $\Delta t$, and all distances are expressed in units of $a$.
The diffusion coefficient is then expressed in units of $a^2/\Delta t$.

To mitigate finite-size artifacts associated with periodic boundaries, we have limited the persistence time $\tau_p$ so that particles do not travel distances comparable to the system linear size without reorienting.
Without this constraint, particles could traverse the entire domain, wrap around the periodic box, and artificially interact with their own periodic images (particularly if $\tau_h$ is large), thereby generating spurious dynamical correlations.
%Under this condition, directed motion and random reorientation occur on physically meaningful length scales.
Hence, we have varied the persistence time logarithmically within the interval $\tau_p \in [1,\,1000]$.
In turn, we have varied the healing time (which does not introduce further comparable finite-size artifacts) logarithmically within the interval $\tau_h \in [1,320]$.

For clarity, a summary of the parameter ranges explored in the simulations is provided in Table~\ref{tab:parameters} at the end of this section.

\begin{table}[t]
\centering
\caption{Summary of the parameter ranges explored in the simulations.}
\begin{tabular}{lcc}
\hline
Parameter & Notation & Range \\
\hline
Lattice size & $L$ & $1024$ \\
Packing fraction & $\phi$ & $0.016 \leq \phi \leq 0.256$ \\
Persistence time & $\tau_p$ & $1 \leq \tau_p \leq 10^{3}$ \\
Healing time & $\tau_h$ & $1 \leq \tau_h \leq 320$ \\
Time step & $\Delta t$ & $1$ \\
Lattice constant & $a$ & $1$ \\
\hline
\end{tabular}
\label{tab:parameters}
\end{table}

\section{Results}
\label{sec:Results}

\subsection{Fixed network topology (\texorpdfstring{$\tau_h=1$}{th=1})}
\label{subsec:topFixed}
We first consider the case in which the network topology is fixed, with all bonds open throughout the simulation.
This case can be more conveniently implemented by setting $\tau_h=1$, so that closed bonds are healed (re-opened) after each time step.
The results of this section will provide a check, as there is a vast literature describing the dynamical behavior of run-and-tumble particles in two dimensions, both in the continuum~\cite{MartensBocquet2012} and on discrete lattices~\cite{SotoGolestanian2014}.
At fixed packing fraction, the effective diffusion coefficient is expected to linearly increase with $\tau_p$ at low persistence time ($D_{\text{eff}} \propto \tau_p$), reach a maximum (usually associated with the onset of clustering in the system), and finally decrease with $\tau_p$ as $D_{\text{eff}} \propto {\tau_p}^{-1}$ at high persistence time.
(The low-$\tau_p$ result $D_{\text{eff}} \propto \tau_p$ can be surmised from the mean-squared displacement of a single active particle~\cite{villa2020run}; in Appendix~\ref{app:PowerLaws}, we present theoretical arguments for $D_{\text{eff}} \propto {\tau_p}^{-1}$ at high-$\tau_p$.)
As we shall discuss in the next paragraph, our results corroborate this picture.
Also, we introduce additional measurements that shed light into the relation between clustering and optimal diffusion in these systems. 

\begin{figure}
\begin{center}
  \includegraphics[width=0.8\columnwidth]{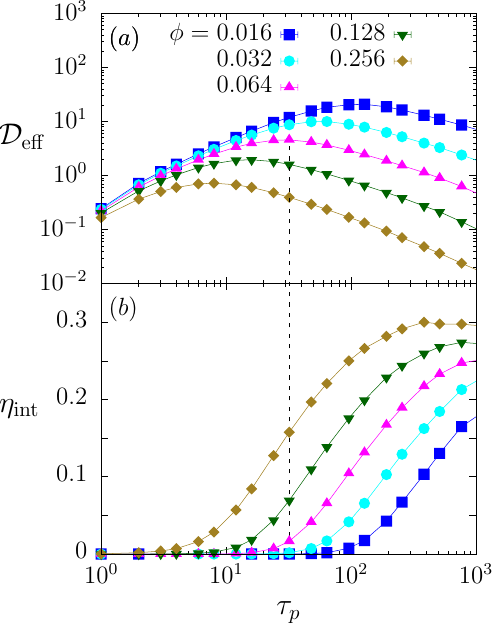}
\end{center}
  \caption{
(a) Effective diffusion coefficient as a function of persistence time for different packing fractions.
(b) Fraction of particles in the ``deep interior'' of a cluster (particles surrounded by $z$ other particles), with respect to the total number of particles with at least one neighbor.
The dashed vertical line indicates $\tau_{p}^{\star}\approx 32$, corresponding to the maximum diffusion for the system with packing fraction $\phi = 0.064$ (magenta curve), as well as the onset of a finite population of particles in the interior of a cluster.
}
  \label{fig:fixedTopologyDeta}
\end{figure}

Figure~\ref{fig:fixedTopologyDeta}a shows the effective diffusion coefficient $D_{\text{eff}}$ as a function of persistence time $\tau_{p}$, for different packing fractions $\phi$. Error bars are smaller than the marker size in all plots shown in this paper.
As anticipated, at fixed $\phi$, $D_{\text{eff}}$ linearly increases with $\tau_p$ at low $\tau_p$, reaches a maximum at some optimal persistence time ${\tau_p}^*$, and then decreases as ${\tau_p}^{-1}$ at high $\tau_p$.
This global behavior is well reproduced for a wide range of values of $\phi$.
As expected, an increase in $\phi$ leads to an overall decrease of $D_\text{eff}$, due to the repulsive interactions between particles.
Besides, an increase in $\phi$ leads to a decrease of the optimal persistence time ${\tau_p}^*$ --- higher packing fraction facilitates agglomeration, which is ultimately responsible for the decrease of $D_\text{eff}$ at high $\tau_p$.

We now seek to quantify agglomeration and clarify its connection to optimal persistence.
Doing so will prove relevant even in the absence of true MIPS, which is usually the case for excluded-volume models without translational noise~\cite{hawthorne2025role,partridge2019critical}.
In fact, for lattice models of persistent particles that interact via excluded volume, there is a simple crossover from weak to strong agglomeration as persistence increases, without signatures of a sharp transition~\cite{soto2014run}.
For reasons that shall become clear, we calculate the fraction of particles located in the ``deep interior'' of a cluster, with respect to the total number of particles with at least one neighbor.
This is defined as
\begin{equation}
\label{eq:1}
\eta_\text{int} = \dfrac{\sum_{\alpha}\delta_{N_{\alpha},z}}{\sum_{\alpha}\Theta(N_{\alpha})},\qquad N_{\alpha}=\sum_{\ell=0}^{z-1}\sigma_{\varphi(x_{\alpha,\ell})}.
\end{equation}
Here $N_{\alpha}$ is the number of occupied nearest-neighbors of particle $\alpha$, $\delta_{i,j}$ is the Kronecker delta, and $\Theta$ is the Heaviside function.
For our triangular lattice model, $\eta_\text{int}$ is just the number of particles with six neighbors divided by the number particles with at least one neighbor. For the densities considered in this work, the denominator in Eq.~(\ref{eq:1}) is always nonzero, so $\eta_{\mathrm{int}}$ is computed over all Monte Carlo frames without ambiguity.
Our definition of $\eta_\text{int}$ is motivated as follows.
We observed that, to enter the regime where diffusion decreases with persistence due to clustering, it is not sufficient that more particles get blocked as persistence is increased.
Temporarily blocked particles can remain significantly mobile on average when residing in small clusters, as particles near cluster interfaces escape more frequently.
Diffusion begins to decrease with persistence only when clusters are large enough to contain many particles in their deep interior.

This is exactly what we verify by plotting $\eta_\text{int}$. Figure~\ref{fig:fixedTopologyDeta}b shows that ${\tau_p}^*$ correlates very well with the value of $\tau_p$ marking the onset of appreciable $\eta_\text{int}$.
For instance, see the black dashed line connecting figures~\ref{fig:fixedTopologyDeta}a and~\ref{fig:fixedTopologyDeta}b: The optimal persistence time marks both the maximum of $D_\text{eff}$ and (approximately) the onset of $\eta_\text{int}$ for $\phi=0.064$ (magenta triangles in the figure).
Although the crossover to finite $\eta_\text{int}$ is not sharp, this result provides evidence for agglomeration as the mechanism responsible for the non-monotonic diffusion in these systems.

\begin{figure}[!ht]
  \includegraphics[width=\columnwidth]{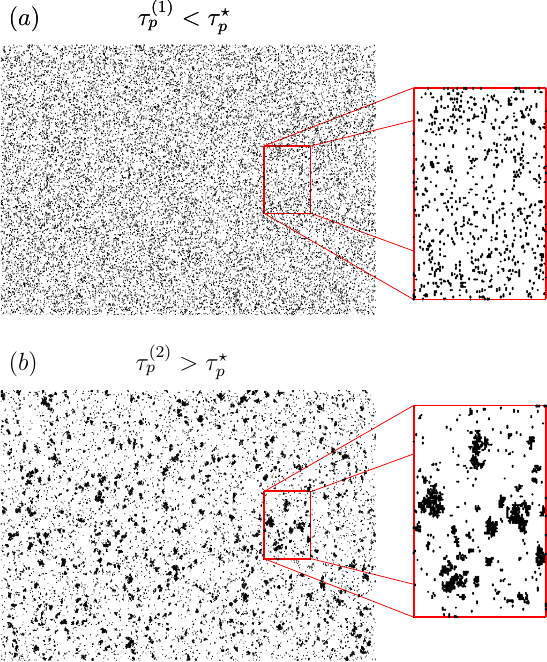}
  \caption{
(a,b) Snapshots corresponding to a single realization of the steady state for $\phi = 0.064$ at persistence times $\tau_{p}^{(1)}=10$ (a) and $\tau_{p}^{(2)}=100$ (b), respectively.
}
  \label{fig:Snapshots}
\end{figure}

To illustrate the clustering mechanism responsible for optimal persistence and the subsequent decrease of diffusion at larger $\tau_p$, Figs.~\ref{fig:Snapshots}a and~\ref{fig:Snapshots}b show snapshots of particle configurations in the lattice for $\phi=0.064$, and $\tau_p=\tau_p^{(1)} =10<{\tau_p}^*$ (a), and $\tau_p=\tau_p^{(2)}=100>{\tau_p}^*$ (b).
These snapshots show clearly that although there are small clusters already present in (a) $\tau_p<{\tau_p}^*$, only in (b) ($\tau_p>{\tau_p}^*$) one can see the formation of large clustered structures with an appreciable fraction of particles in the deep interior of a cluster. That said, we highlight once again that there is no sharp clustering transition here, only an optimum in diffusivity due to increased agglomeration.

\begin{figure}
  \includegraphics[width=\columnwidth]{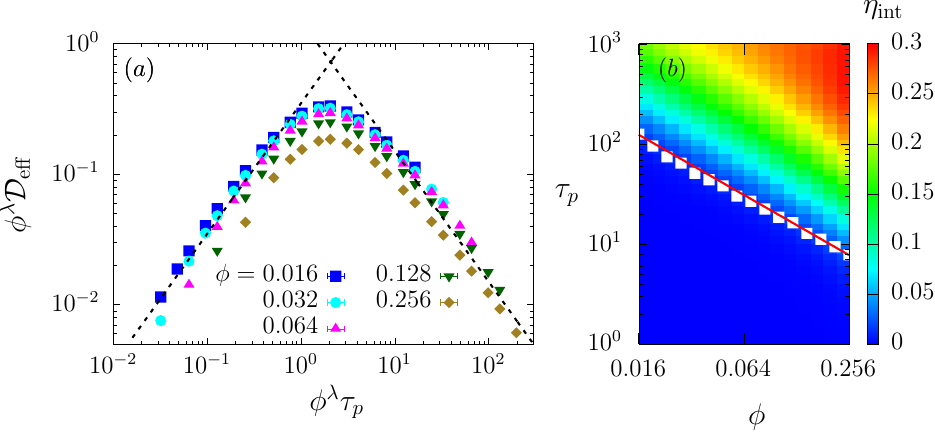}
  \caption{
(a) Scaling collapse plot showing re-scaled effective diffusion coefficient as a function of re-scaled persistence time.
The exponent $\lambda \approx 1$ was chosen to yield the best scaling of the data.
(b) Density plot of the fraction of particles in the deep interior of a cluster as a function of packing fraction and persistence time.
White squares represent a numerical calculation of the optimal persistence time at each packing fraction, and the solid red line is a best fit using Eq.~\eqref{eq:tauStar}.
}
  \label{fig:Scaling}
\end{figure}

We end this section with a brief discussion about the apparent invariant scaling behavior depicted in Fig.~\ref{fig:fixedTopologyDeta}a.
Notice that all curves seem to have the same shape in a log-log scale, which suggests that $D_{\text{eff}}$ follows a simple scaling form,
\begin{equation}
D_{\text{eff}} = \phi^{-\lambda} \mathcal{F} \left( \tau_p \, \phi^\lambda \right),
\label{eq:ScalingDeff}
\end{equation}
where $\lambda$ is a scaling exponent, and $\mathcal{F}$ is a scaling function.
The fact that the scaling exponents for $D_\text{eff}$ and $\tau_p$ are the same is a consequence of the linear behavior in the low-$\tau_p$ regime.
In fact, at low $\tau_p$, Eq.~\eqref{eq:ScalingDeff} implies that $\mathcal{F}(x)\sim x$ at low $x$, so that $D_\text{eff} \sim \tau_p$ independent of $\phi$.
In turn, at high $\tau_p$ we must have $\mathcal{F}(x)\sim x^{-1}$, so that we can derive a simple estimate for ${\tau_p}^*$ from the condition that the two regimes meet at ${\tau_p}^*$:
\begin{equation}
C_- \, \phi^{-\lambda} {\tau_p}^* \, \phi^\lambda = C_+ \, \phi^{-\lambda} \left({\tau_p}^* \, \phi^\lambda \right)^{-1} \rightarrow {\tau_p}^* \sim \phi^{-\lambda},
\label{eq:tauStar}
\end{equation}
where $C_{-}$ and $C_{+}$ are the amplitudes of the power laws for $D_\text{eff}$ below and above the optimal persistence, respectively.
Figure~\ref{fig:Scaling}a shows a scaling collapse plot for re-scaled diffusion coefficient $D_\text{eff} \phi^\lambda$ as a function of rescaled persistence time $\tau_p \, \phi^\lambda$.
The exponent $\lambda \approx 1$ is chosen to yield the best collapse of the data. (Curiously, this exact same result is obtained for the agglomeration transition line in other models through a kinetic balance theory~\cite{redner2013,de2025quorum}.)
Notice that the scaling collapse becomes worse at higher values of packing fraction, with $\phi=0.256$ showing the most pronounced deviation in Fig.~\ref{fig:Scaling}a. This deviation of the proposed scaling at large densities can be partly due to the significant changes both in spatial cluster structures and in the role of interactions for this regime.
The incorporation of suitable corrections to scaling~\cite{cardy1996scaling, aharony1980university, aharony1983nonlinear, raju2019normal, thornton2025universal} could potentially lead to an extended range of densities for the scaling regime, but is beyond the scope of the present work.
Figure~\ref{fig:Scaling}b shows a density plot of $\eta_\text{int}$ as a function of packing fraction $\phi$ and persistence time $\tau_p$.
The white squares show the simulation results for ${\tau_p}^* = {\tau_p}^* (\phi)$ and solid red line shows the best fit using our theoretical prediction for ${\tau_p}^* \sim \phi^{-\lambda}$.
Notice that there is excellent agreement between the direct calculation of ${\tau_p}^*$, the onset of non-zero $\eta_\text{int}$ and our theoretical prediction in Eq.~\eqref{eq:tauStar}.

\subsection{Adjustable network topology (\texorpdfstring{$\tau_h>1$}{th>1})}
\label{subsec:AdjustableNetworks}

We now consider $\tau_h>1$, i.e.\ the network topology changes due to particle motion.
Now the healing time $\tau_h$ --- the average time it takes for a closed bond to re-open --- plays a fundamental role.
The larger the value of $\tau_h$, the larger the chance that a particle will be blocked by a closed track.
Hence one would naively expect that the effect of closed tracks is in a sense similar to steric interactions due to other particles, and that dependence of transport on $\tau_h$ should be similar to the dependence on $\phi$.
We will show that this is not always the case: Unlike the packing fraction case, the optimal persistence time ${\tau_p}^*$ increases with $\tau_h$.

\begin{figure}
  \includegraphics[width=\columnwidth]{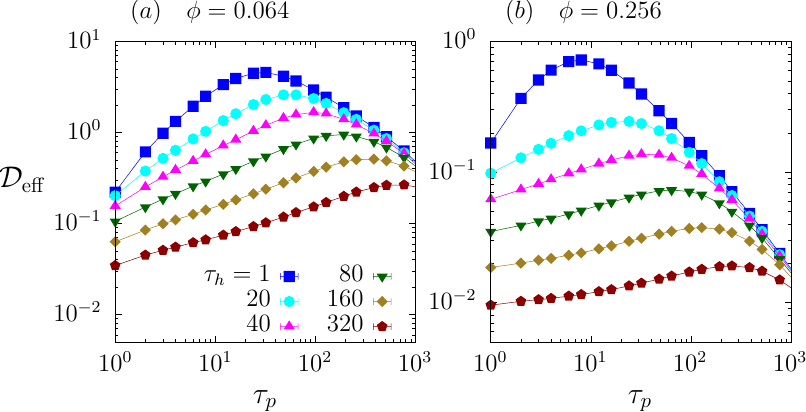} \\
  \caption{
Effective diffusion coefficient as a function of persistence time for packing fractions $\phi = 0.064$~(a) and $\phi = 0.256$~(b), and several values of the healing time $\tau_h$. 
}
  \label{fig:DiffusionFiniteTh}
\end{figure}

Figures~\ref{fig:DiffusionFiniteTh}a and ~\ref{fig:DiffusionFiniteTh}b show the effective diffusion coefficient as a function of persistence time for several values of $\tau_h$ and fixed packing fractions $\phi=0.064$, and $\phi=0.256$, respectively.

As expected, we observe an overall decrease of $D_\text{eff}$ with the increase of $\tau_h$.
Also, we observe similar qualitative behavior of each $\tau_p \times D_\text{eff}$ curve for fixed $\tau_h$ in Fig.~\ref{fig:DiffusionFiniteTh}a-b and for fixed $\phi$ in Fig.~\ref{fig:fixedTopologyDeta}a.
There is an initial increase of $D_\text{eff}$ with $\tau_p$ at low $\tau_p$, which now shows deviations from linear behavior at both high $\tau_h$ and high $\phi$.
At intermediate values of $\tau_p$, $D_\text{eff}$ reaches a maximum, and then starts decreasing as ${\tau_p}^{-1}$ at high $\tau_p\gg \tau_h$.
In fact, we will check that the dominant blocking mechanism for a particle in the regime $\tau_p \gg \tau_h$ is the steric repulsion due to agglomeration, which explains why $D_\text{eff}$ does not depend on $\tau_h$ in this regime.
More surprisingly, in contrast with the behavior we have described for fixed topology and varying packing fraction~\cite{LevisBerthier2014}, the optimal persistence time ${\tau_p}^*$ increases with $\tau_h$.
To explain this counterintuitive behavior, we need to take a closer look into the two different mechanisms for particle blocking (the diffusion brakes) in this model. 

First, let us note that unlike the fixed topology case ($\tau_h=1$), for $\tau_h>1$ there are two blocking mechanisms: (i) Blocking by particles, and (ii) blocking by closed tracks.
Crucially, mechanisms (i) and (ii) lead to different behaviors upon increasing $\tau_p$.
In Section~\ref{subsec:topFixed} we have described mechanism (i) in detail.
We have argued that the non-monotonic dependence of $D_\text{eff}$ on $\tau_p$ is largely connected with the onset of a finite population of particles in the deep interior of clusters.
Within the low-persistence regime, increasing $\tau_p$ makes active particles diffuse faster; in the high-persistence regime, diffusion becomes slower due to agglomeration---particles in the deep interior of a cluster will be more strongly trapped the larger the $\tau_p$.
This does not happen for mechanism (ii).
As we show in Appendix~\ref{app:NoParticleInt}, if we turn off steric repulsive interactions (by allowing more than one particle per site), we observe that the effective diffusion coefficient always increases with persistence time.
Upon increasing $\tau_h$, it is the redistribution of blocking mechanisms that is ultimately responsible for the unexpected increase of ${\tau_p}^*$ with healing time.

\begin{figure}
  \includegraphics[width=\columnwidth]{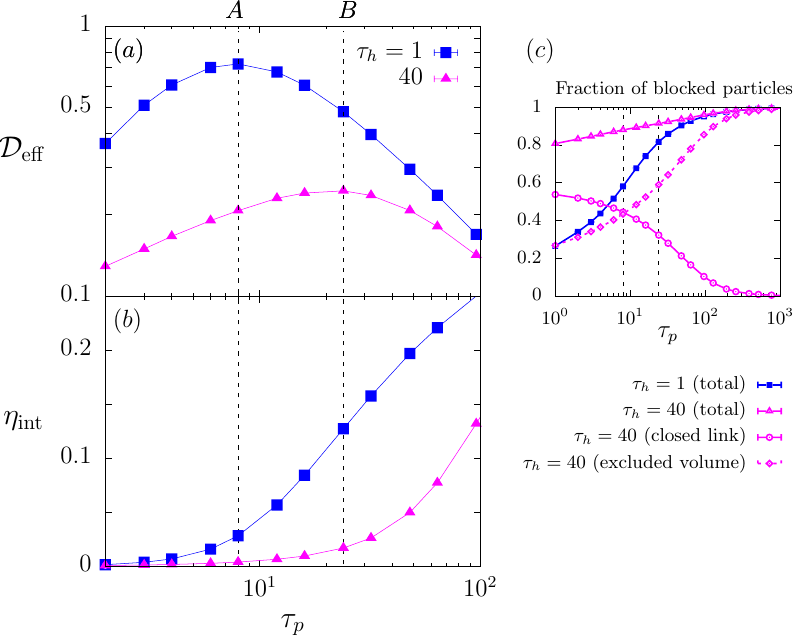}
  \caption{
(a)-(b): Effective diffusion coefficient (a) and fraction of particles in the deep interior of a cluster (b) as a function of persistence time for packing fraction $\phi=0.256$, and healing times $\tau_h=1$ and $\tau_h=40$.
The dashed lines $A$ and $B$ indicate the optimal persistence time for each healing time.
(c) Fraction of blocked particles as a function of the persistence time.
}
  \label{fig:Mechanisms}
\end{figure}

Figure~\ref{fig:Mechanisms}a shows the effective diffusion coefficient for fixed density $\phi=0.256$, and healing times $\tau_h=1$ (no closed bonds) and $\tau_h=40$ (with closed bonds).
Let us consider $\tau_p=\tau_p^{(A)}$, i.e.\ line $A$ corresponding to ${\tau_p}^*$ when $\tau_h=1$. 
As expected, $D_\text{eff}$ decreases with $\tau_h$ due to the introduction of blocking mechanism (ii).
In turn, the fraction of particles in the deep interior of a cluster $\eta_\text{int}$ \emph{also decreases} as $\tau_h$ is increased, as show in Fig.~\ref{fig:Mechanisms}b.
Since we have observed that it is this population that is responsible for reducing diffusion at large persistence times [as $D_\text{eff}$ always increases with $\tau_p$ for mechanism (ii)], the optimal persistence time for $\tau_h>1$ has to be located at a value $\tau_h^{(B)}>\tau_h^{(A)}$, thus explaining the observed increase of the optimal persistence time with healing time.

At sufficiently large persistence time, particles organize themselves into larger clusters, and mechanism (i) becomes dominant. 
Figure~\ref{fig:Mechanisms}c shows the fraction of blocked particles as a function of persistence time for $\tau_h=1$ and $\tau_h=40$, with respect to the total number of particles.
For $\tau_h=1$ there is only one mechanism, whereas for $\tau_h=40$ we show the fraction of blocked particles for each mechanism as well as the total fraction.
The dashed lines corresponds to lines $A$ and $B$ in Fig.~\ref{fig:Mechanisms}a-b.
Notice that for large $\tau_p$, almost all blocked particles are due to excluded volume interactions.
This explains how the effective diffusion coefficient becomes independent of healing time at large persistence time.

\begin{figure}[!h]
  \includegraphics[width=\columnwidth]{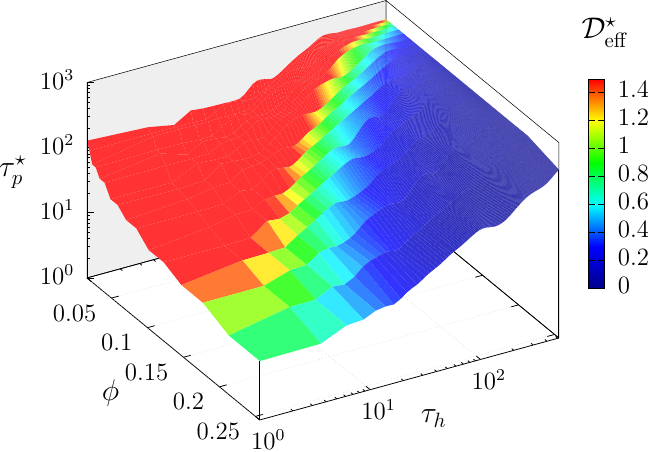}
  \caption{Optimal persistence time as a function of packing fraction and `healing' time, separating regimes where diffusion increases (below $\tau_p$) and decreases (above $\tau_p$) with persistence time.
  The color in the surface corresponds to the value of the effective diffusion coefficient at the optimum.}
  \label{fig:3DPlot}
\end{figure}

The main results of this section are summarized in the three-dimensional plot shown in Fig.~\ref{fig:3DPlot}, in terms of optimal persistence time, packing fraction and healing time.
For points below and above the surface, the effective diffusion coefficient increases and decreases with persistence time, respectively.
Although the optimal persistence time decreases with packing fraction, it increases with healing time, due to the different blocking mechanisms associated with excluded volume and closed tracks.
The surface is colored according to the value of the effective diffusion constant at the optimum for each packing fraction and healing time.

Finally, we note that unlike the fixed-topology case ($\tau_h = 1$), we do not observe signatures of invariant scaling behavior in the dependence of $D_\textrm{eff}$ on $\tau_h$.
An inspection of Fig.~\ref{fig:DiffusionFiniteTh} shows that the shape of the $\tau_p \times D_\textrm{eff}$ curves changes systematically on a log--log scale as $\tau_h$ varies.
Overall, the transport properties are governed by a nontrivial interplay between $\tau_h$ and $\tau_p$, which influences both the onset and release of blocking, as well as the relative contribution of the different blocking mechanisms.

\section{Conclusions}
\label{sec:Conclusions}

We now make some concluding remarks about our paper.
Here we have introduced and investigated transport properties of a model of active particles moving in an adjustable network.
In our model, particles move along bonds connecting nearest-neighbor sites of a network, and execute a traditional discrete-version of run-and-tumble motion characterized by a persistence time $\tau_p$.
As each particle moves, it leaves a trail of closed links that cannot be used by itself nor by other particles.
This trail re-opens, or heals, after a characteristic healing time $\tau_h$.
Particles also interact directly via excluded-volume interactions.
We find that particle-particle interactions (particles blocked by excluded volume) and particle-network interactions (particles blocked by closed tracks) lead to different diffusive behavior as persistence time is varied. 
In the limit of no network topology change (i.e.\ no closed bonds), we recover the expected non-monotonic dependence of the diffusion coefficient on persistence time, with a characteristic increase and decrease at low and high persistencies, respectively.
We also use simple scaling arguments to describe these regimes.
When we allow network topology changes ($\tau_h>1$ in our units), we find that the two blocking mechanisms (particle-particle and particle-network interactions) lead to different behaviors for optimal diffusion in these systems.
As a result, the optimal persistence time decreases with packing fraction, but increases with healing time. We expect either our model or simple variants of it to become relevant in the description of transport properties of a broad variety of systems of active matter moving in adjustable media, such as self-propelled swimmers in microfluidic devices, slime mold Physarum polycephalum and fibroblasts on cartilage tissues.

\appendix
\section{Statistical convergence}
\label{app:StatConv}
In this appendix, we provide additional details about our numerical simulations, and the procedures we use to ensure equilibration and statistical convergence of the observables.
Notice that standard arguments for Monte Carlo convergence do not necessarily apply for out-of-equilibrium systems.
Hence, we have taken special care to determine the number of simulation steps required for the system to reach a stationary state.

Our approach relies on a careful analysis of the mean square displacement (MSD) of the active particles.
First, we let the system evolve for $N_{\mathrm{ther}} = 10^{5}$ Monte Carlo Steps $(\mathrm{MCS})$.
Then, we measure the MSD as a function of time for additional $N_{\mathrm{step}} = 10^{5}\;\mathrm{MCS}$.

\begin{figure}
  \includegraphics[width=\columnwidth]{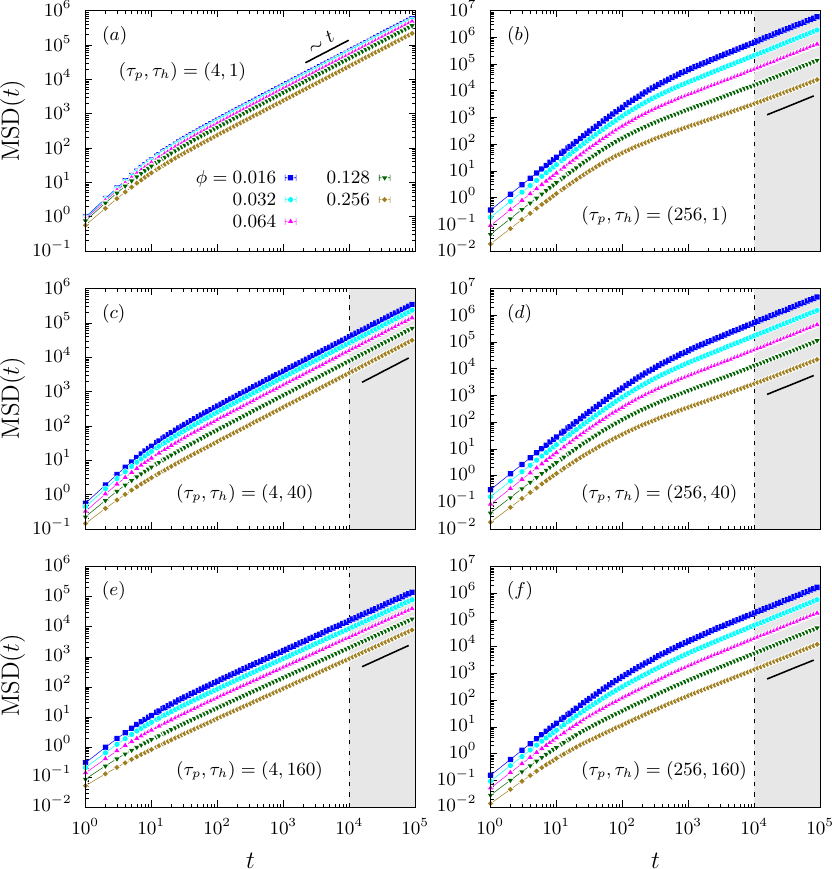}
 \caption{Time evolution of the mean square displacement for several values of packing fraction $\phi$, and a few values of persistence time $\tau_p$ and healing time $\tau_h$.
 The observables reported in this paper (e.g. $D_{\mathrm{eff}}$ and $\eta_{\mathrm{int}}$) are computed from time averages within the gray regions ($10^4\leq t\leq 10^5\;\mathrm{MCS}$), which are deep inside the diffusive regime ($\mathrm{MSD}(t)\sim t$).
 The solid black line represents linear behavior.}
  \label{fig8}
\end{figure}

Figures~\ref{fig8}a--f show the MSD as a function of time for different values of the packing fraction, persistence time, and healing time.
The system exhibits a clear diffusive regime ($\mathrm{MSD}(t) \sim t$) for $t \gtrsim 10^{4}$ MCS (see the curves within the gray regions in the panels).
We compute the reported observables $D_{\mathrm{eff}}$ and $\eta_{\mathrm{int}}$ for times within this time window ($10^{4} \leq t \leq 10^{5}\,\mathrm{MCS})$.

For each set of parameters, we perform $N_{\mathrm{real}}$ independent realizations, with $N_{\mathrm{real}} = 10$ for $\phi \leq 0.064$ and $N_{\mathrm{real}} = 5$ otherwise.
For each realization, we initialize the system using different configurations and random number seeds to ensure statistical independence.
In our model, disorder is not quenched but emerges dynamically from the particle activity, leading to distinct microscopic configurations across realizations.
We compute the reported observables for each realization and subsequently take an average of the results.
Statistical uncertainties are estimated as the standard error of the mean across realizations.
In all plots shown in this paper, these uncertainties are smaller than the marker size used in the figure plots.

\begin{figure}
  \includegraphics[width=\columnwidth]{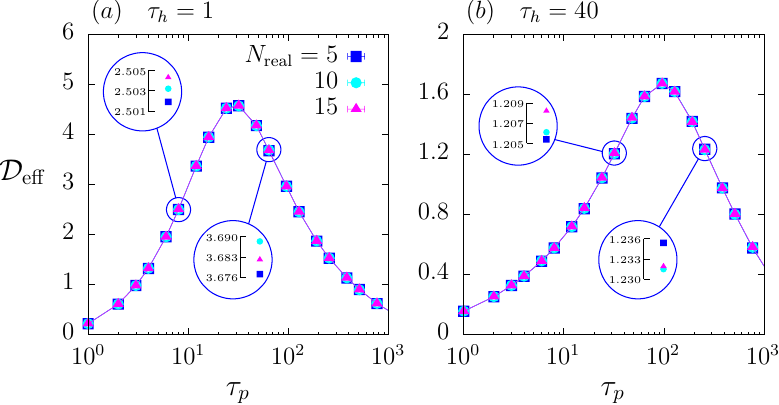}
 \caption{Effective diffusion coefficient as a function of the persistence time for different numbers of realizations ($N_{\mathrm{real}}=5,10,15$), at fixed packing fraction $\phi=0.064$ and healing times $\tau_{h}=1$ (a) and $\tau_{h}=40$ (b).}
  \label{fig9}
\end{figure}

Figure~\ref{fig9}a shows $D_{\mathrm{eff}}$ as a function of $\tau_{p}$ for different numbers of realizations ($N_{\mathrm{real}}=5,10,15$), at fixed packing fraction $\phi=0.064$ and $\tau_{h}=1$.
The results show negligible variation, indicating that statistical convergence is already achieved for $N_{\mathrm{real}}=5$.
For instance, in the zoomed-in region around $\tau_{p}=8$ (upper inset), indicate that variations in the mean value of $D_{\mathrm{eff}}$ are of order $10^{-3}$ in this scale, corresponding to a negligible relative variation $< 0.1\%$.
Similarly, in the lower inset at $\tau_{p}=64$, variations in the mean value of $D_{\mathrm{eff}}$ are of order $10^{-2}$ in this scale, corresponding to a relative variation $< 1\%$.
The same analysis is shown in Fig.~\ref{fig9}b, for an adjustable network with $\tau_{h}=40$ and $\phi=0.064$.

\section{High-persistence behavior for fixed network topology ($\tau_h=1$)}
\label{app:PowerLaws}

We now sketch a previously unreported theoretical argument for the result $D_\text{eff} \propto \tau_p^{-1}$ for high $\tau_p$ and $\tau_h=1$.
Although this is not a rigorous derivation, it sheds light into the quantitative origin of the result.
We proceed by analogy with a different scenario.
In the the run-stop-tumble (RST) model~\cite{curatolo2020cooperative}, particles run ballistically and then, at rate $\beta_\text{in}$, enter into a stopped mode.
Then, at rate $\beta_\text{out}$, they leave the stopped mode and return to the run mode with a new director.
That is, the stopped mode is a prolonged tumbling event during which the particle does not move. The authors present an expression for the diffusivity of these particles:
\begin{equation}
D_\text{eff}^\text{RST} \propto \frac{\beta_\text{out}}{\beta_\text{in}(\beta_\text{in}+\beta_\text{out})}.
\end{equation}

Now we introduce the analogy.
For simple run-and-tumble particles with excluded volume, in the high-$\tau_p$ regime, mobile particles can be expected to move ballistically between clusters and remain stopped or blocked at their interfaces for some time of order $1/\tau_p$.
This is similar to the RST scenario.
The time between two stops is assumed to be proportional to the distance between clusters, which we assume to be proportional to $\sqrt{\tau_p}$, based on previous results for the cluster sizes on square lattices~\cite{soto2014run}.
Therefore, we can assume $\beta_\text{in}\propto \sqrt{\tau_p}$ and $\beta_\text{out}\propto \tau_p^{-1}$.
Finally, only mobile particles contribute to diffusivity.
The fraction of mobile particles depends on $\tau_p$ and follows an empirical expression that was also previously obtained for square lattices~\cite{soto2014run}.
The expression reads $\alpha_p(1-\phi)/[\alpha_p + 0.6(1-\alpha_p)\phi]$, into which we insert $\alpha_p=\tau_p^{-1}$.
Multiplying the expression for $D_\text{eff}^\text{RST}$ above by this fraction of mobile particles and expanding around $\tau_p^{-1}=0$ gives, to first order, $D_\text{eff}\propto\tau_p^{-1}$.
Incidentally, expanding around $\tau_p=1$ gives the low-$\tau_p$ result, $D_\text{eff} \propto \tau_p$.

\section{Model without steric interactions}
\label{app:NoParticleInt}
To isolate the effect of network topological rearrangements on particle transport, we consider a simplified model with no steric interactions between particles. This allows us to suppress the clustering-induced blocking mechanism and focus exclusively on the impact of closed tracks on transport properties of our model.

Except for a few important differences, the model is essentially the same as the one described in Sec.~\ref{sec:Model}.
Now the occupation variable of lattice site $i$ has to allow for multiple particles, so that $\sigma_i\in\{0,\dots,N_{p}\}$. Coordinates $x_{\alpha}$ for particles $\alpha$ are no longer constrained to be distinct.
All the remaining degrees of freedom --- namely the bond variables $\kappa_{i\ell}$ and the self-propulsion directions $n_{\alpha}$ --- are left unchanged. 
The dynamical rules are modified accordingly. The run-and-tumble dynamics is now free from steric blocking, while the bond-closing and healing processes are implemented exactly as in the original model.

\begin{figure}
  \includegraphics[width=\columnwidth]{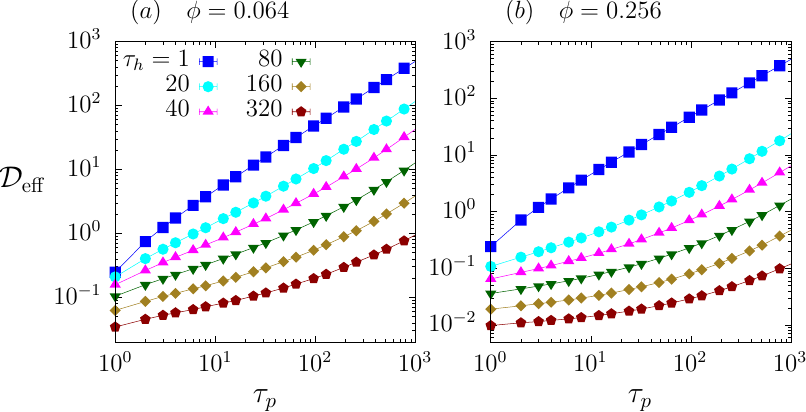}
 \caption{ Effective diffusion coefficient as a function of the persistence time for different healing times, at packing fractions $\phi = 0.064$ (a) and $\phi = 0.256$ (b).
 In this model, steric interactions are absent.}
  \label{fig4}
\end{figure}

In this simplified setting, where blocking due to particle-particle interactions is absent, we find that the blocking mechanism associated with closed bonds does not give rise to a non-monotonic dependence on $\tau_{p}$.
Instead, we find that increasing $\tau_{p}$ always enhances transport, and $D_{\text{eff}}$ increases monotonically for all values of $\tau_{h}$.
We demonstrate this behavior in Fig.~\ref{fig4}a and~\ref{fig4}b, which show the effective diffusion coefficient is shown as a function of $\tau_{p}$ for different values of $\tau_{h}$ and fixed packing fractions $\phi=0.016$ (a) and $\phi=0.256$ (b), respectively.

\section*{Author contributions}
All authors contributed equally to: Conceptualization, Formal analysis, Investigation, Methodology, Validation and Writing -- review \& editing.
W.G.C.O.: Data curation, Software, Visualization and Writing -- original draft.
P.d.C.: Data curation, Software and Visualization.
H.L.: Funding acquisition and Supervision.
D.B.L.: Funding acquisition, Project administration, Resources, Supervision, Visualization and Writing -- original draft. 

\section*{Conflicts of interest}
There are no conflicts to declare.

\section*{Data availability}
The simulation code used in this study is publicly available at Zenodo~\cite{oropesa2026active}. The repository is also maintained at 
\url{https://github.com/Liarte-Group/Active-Matter-Adjustable-Networks}. 
All simulation datasets are available at the same Zenodo record.

\section*{Acknowledgements}
%The authors acknowledge the support of ICTP-SAIFR (São Paulo, Brazil), funded by the São Paulo Research Foundation (FAPESP) under Grant No.\ 2021/14335-0.
The authors thank Prof.\ Andr\'e Vieira for providing access to his GPU computing resources.
The authors also thank Profs.\ Itai Cohen, James Sethna, Jennifer Schwartz, M.\ Lisa Manning and Rodrigo Soto for valuable discussions.
W.G.C.O.\ acknowledges support from FAPESP under Grant No.\ 2024/23876-3.
P.d.C.\ acknowledges support from FAPESP under scholarships 2021/10139-2 and 2022/13872-5, and from CNPq under Conhecimento Brasil grant 446379/2024-7 and scholarship 316196/2025-8.
H.L.\ acknowledges support from the German Research Foundation DFG within the project LO418/29-1.
D.B.L.\ acknowledges support from FAPESP under Grants No.\ 2021/14285-3, No.\ 2022/09615-7 and No.\ 2023/14815-8.

%%%END OF MAIN TEXT%%%

%The \balance command can be used to balance the columns on the final page if desired. It should be placed anywhere within the first column of the last page.

\balance

%If notes are included in your references you can change the title from 'References' to 'Notes and references' using the following command:
%\renewcommand\refname{Notes and references}

%%%REFERENCES%%%

\bibliography{rsc} %You need to replace "rsc" on this line with the name of your .bib file
\bibliographystyle{rsc} %the RSC's .bst file

\end{document}